\documentclass{revtex4-2}

\usepackage{graphicx}
\usepackage{dcolumn}
\usepackage{bm}

\usepackage[utf8]{inputenc}
\usepackage[T1]{fontenc}
\usepackage{mathptmx} 

\usepackage{ucs}
\usepackage[utf8]{inputenc}
\usepackage{amsmath}
\usepackage{float}
\newcommand{\Tr}{\text{Tr}}

\begin{document}
\title{Bogomol'nyi-like Equations in Gravity Theories}

\author{Ardian Nata Atmaja} 
 \email[Corresponding author: ]{ardi002@brin.go.id}
\affiliation{
  Research Center for Quantum Physics, National Research and Innovation of Republic of Indonesia (BRIN),\\
  Komplek PUSPIPTEK, Serpong, Tangerang Selatan 15310, Indonesia.
}

\date{\today} 

\begin{abstract}
Using the BPS Lagrangian method, we show that gravity theory coupled to matter in various dimensions may possess Bogomol'nyi-like equations, which are first-order differential equations, satisfying the Einstein equations and the Euler-Lagrange equations of classical fields ($U(1)$ gauge and scalar fields). In particular we consider static and spherically symmetric solutions by taking proper ansatzes and then we find an effective Lagrangian density that can reproduce the Einstein equations and the Euler-Lagrange equations of the classical fields. We consider the BPS Lagrangian density to be linear function of first-order derivative of all the fields. From these two Lagrangian desities we are able to obtain the Bogomol'nyi-like equations whose some of solutions are well-known such as Schwarzschild, Reissner-Nordstr\"{o}m, Tangherlini black holes, and the recent black holes with scalar hair in three dimensions [Phys. Rev. D 107, 124047]. Using these Bogomol'nyi-like equations, we are also able to find new solutions for scalar hair black holes in three and four dimensional spacetime. Furthermore we show that the BPS Lagrangian method can provide a simple alternative proof of black holes uniqueness theorems in any dimension.
\end{abstract}

\maketitle

\tableofcontents

\section{Introduction}
In nonlinear field theory, Bogomol'nyi equations are closely related to attempts on finding exact solutions of the theory and their stabilities~\cite{Bogomolny:1975de}. One of the earliest attempts in finding exact solutions were done by Prasad and Sommerfield in the case of $SU(2)$ Yang-Mills-Higgs model~\cite{Prasad:1975kr}. In particular they found exact solutions of 't Hooft-Polyakov monopoles and Julia-Zee dyons in the limit where both mass and quartic-coupling interaction of the Higgs fields go to zero. Bogomol'nyi then found by rewriting the energy functional into completed square form, so called Bogomol'nyi trick or decomposition, those exact solutions minimize the energy functional and are also solutions to first-order differential equations known as Bogomol'nyi equations. Furthermore he found that the total energy of these solutions are propotional to topological charge, and thus are stable, see~\cite{Manton:2004tk,Weinberg:2012pjx} for more detail calculations. These Bogomolny equations turn out to be usefull for studying stability of solutions in the classical field theory and one may be able to find some exact solutions since the problem of solving second-order differential equations, or Euler-Lagrange equations, now reduce to problem of solving these first-order differential equations. In supersymmetric theories, with central charge, the Bogomol'nyi equations can be obtained from variation of fermionic fields, that breaks some of supersymmetric charges and in this context the Bogomol'nyi equations are usually called BPS equations~\cite{Witten:1978mh}.

The Bogomol'nyi trick does not have a systematic procedure, and hence is hard to be employed to more general models. There are several methods have been developed in order to obtain the Bogomol'nyi equations: Strong Necessary Condition~\cite{Sokalski:1979,Sokalski:2001wk}, first-order formalism~\cite{Bazeia:1995en,Bazeia:2005tj,Bazeia:2007df}, {\it On-Shell} method~\cite{Atmaja:2014fha,Atmaja:2015lia}, FOEL (First-Order Euler-Lagrange) formalism~\cite{Adam:2016ipc}, and BPS Lagrangian method~\cite{Atmaja:2015umo}. So far all these methods have been used mostly for (non-supersymmetric) field theories where the space-time metric is flat. Some of these methods have been used to find Bogomol'nyi equations for gravity, but only for a particular case of spatially flat universe metric in the scalar field inflation~\cite{Bazeia:2005tj,Adam:2016ipc}. There are some earliest attempts to find Bogomol'nyi-type equations in the system that include gravity either as a background metric or as dinamical field, such as incorporating the Einstein-Hilbert action,~\cite{Comtet:1979kq,Comtet:1983wt,Forgacs:1994pp,Zhang:2009tt}. Unfortunately there were still no Bogomol'nyi-type equations found satisfying the Einstein equations. However there is an attempt to find Bogomol'nyi equations for black holes by using Bogomol'nyi trick which, as we mentioned before, may not be applicable to more general gravity models~\cite{Miller:2006ay}~\footnote{In there they actually used Bogomol'nyi trick to the Lagrangian density, in similar fashion as the BPS Lagrangian method, and the resulting first-order differential equations were called BPS equations.}. In this article we will try to find first-order differential equations that satisfy the Einstein equations and the Euler-Lagrange equations of classical fields ($U(1)$ gauge and scalar fields) in some of gravity theories and, for an obvious reason, we will be using the BPS Lagrangian method which been applied to many field theories in various dimensions~\cite{Atmaja:2017nlt,Atmaja:2018cod,Atmaja:2018ddi,Atmaja:2019gce,Atmaja:2020iti,Fadhilla:2020rig,PrasetyaTama:2021elj,Fadhilla:2021jiz}. We shall call these first-order differential equations as Bogomol'nyi-like equations since they will be derived from the action, or to be more precise (effective) Lagrangian density, instead of total energy as in the original article by Bogomol'nyi~\cite{Bogomolny:1975de}.

\subsection{BPS Lagrangian Method}

Let us consider an action of $N$ scalar fields $\phi_m\equiv(\phi_1,\ldots,\phi_N)$ in $(d+1)$-dimensions of spacetime,
\begin{equation}
 S=\int d^{d+1}x\sqrt{-g}~\mathcal{L}(\phi_m,\partial_\mu\phi_m),
\end{equation}
with $\partial_\mu\equiv{\partial\over\partial x^\mu}\equiv \left({\partial\over\partial x^0},{\partial\over\partial x^1},\ldots,{\partial\over\partial x^d}\right)$ and $\mu=0,1,\ldots, d$ is the spacetime index. Similar to Bogomol'nyi's trick, we shall rewrite the Lagrangian density into complete squared terms as such
\begin{equation}
 \sqrt{-g}\mathcal{L}\equiv\mathcal{L}_{eff}=\left(\mbox{squared terms}\right)+\mathcal{L}_{BPS},
\end{equation}
where 
\begin{equation}
 \left(\mbox{squared terms}\right)= \sum_{n=1}^N h_n(\phi_m) \left(\partial_\nu \phi_n - f_n(\phi_m,\partial_\mu\phi_m^0) \right)^2,
\end{equation}
with $\partial_\mu\phi_m^0\equiv \{\partial_\mu\phi_m\}-\partial_\nu\phi_n$ for fixed $\nu$ and $n$.
Here $\mathcal{L}_{BPS}$ is defined as BPS Lagrangian density that usually contain only boundary terms by means of its Euler-Lagrange equations are trivially satisfied. Once we determine the form of BPS Lagrangian density then we could find the ``squared terms''. We then define a BPS limit in which $\mathcal{L}_{eff}-\mathcal{L}_{BPS}=0$ and all terms in the ``squared terms" are set to zero as such $\partial_\nu \phi_n = f_n(\phi_m,\partial_\mu\phi_m^0)$ which shall be called Bogomol'nyi equations. Now one may ask what is the explicit form of BPS Lagrangian density?. One of the answer comes from the study of well-known Bogomol'nyi equations in various models using the {\it On-Shell} metode~\cite{Atmaja:2014fha}. For spherically static cases, total energy is proportional to the action, $E=-\int d^{d+1}x~\mathcal{L}_{eff}$, and in the BPS limit the energy of BPS solitons are determined by difference between values of a function $Q\equiv Q(\phi_m)$, there is called BPS energy functional, on the boundary and on the origin, 
\begin{equation}
 E_{BPS}=Q(r\to\infty)-Q(r=0)=\int_{r=0}^\infty dQ=\int dr \left(\sum_{n=1}^N{\partial Q\over\partial\phi_n}\phi_n'(r)\right)=\int dr \mathcal{L}_{BPS}.
\end{equation}
Here $\mathcal{L}_{BPS}$ is a linear function of first-order derivative of fields $\phi_m$ and one can simply show that its Euler-Lagrange equations are indeed trivial. The boundary terms are not restricted only to linear function of $\phi_m'(r)$. In general there are also possible boundary terms contain $\partial_\mu\phi_m$ with power higher than one. As an example for $N=3$ and $d=3$, we may have $\mathcal{L}_{BPS}$ with boundary terms as follows~\cite{Adam:2016ipc}
\begin{equation}
 \mathcal{L}_{BPS}=\sum_{n,o,p} Q^{[ijk]}_{[nop]} \partial_i\phi_n \partial_j\phi_o \partial_k\phi_p,
\end{equation}
where $ Q^{[ijk]}_{[nop]}\equiv  Q^{[ijk]}_{[nop]}(\phi_m)$ is totally antisymmetric tensor in both up and down indices. Here we have used Einstein summation notation over the spatial coordinate indices $i,j,$ and $k$. However $\mathcal{L}_{BPS}$ may also contain non-boundary terms such that its Euler-Lagrange equations,
\begin{equation}
 {\partial\mathcal{L}_{bps}\over\partial \phi_m}-{\partial\over\partial x^\mu}\left({\partial\mathcal{L}_{bps}\over\partial (\partial_\mu\phi_m)}\right)=0,
\end{equation}
are not trivially satisfied and must be considered as additional constraint equations in finding solutions to the Bogomol'nyi equations.

\subsubsection{Case without ansatz}
Let us see how to employ explicitly the BPS Lagrangian method to the $SU(2)$ Yang-Mills-Higgs model with the following action in Minskowski spacetime
\begin{equation}\label{Action YMH}
  S=\int d^4x\sqrt{-g}~ \mathcal{L}=\int d^4x\left(-{1\over 2}\Tr\left(F_{\mu\nu}F^{\mu\nu}\right)+ \Tr\left(D_\mu\Phi D^\mu\Phi\right)-V(|\Phi|)\right),
\end{equation}
with $|\Phi|=2\Tr(\Phi^2),~F_{\mu\nu}=\partial_\mu A_\nu-\partial_\nu A_\mu-ie \left[A_\mu,A_\nu\right],~D_\mu\equiv\partial_\mu-ie\left[A_\mu,\right]$, $A_\mu={1\over 2}\tau^a A^a_\mu,~ \Phi={1\over 2}\tau^a \Phi^a$, with internal index $a=1,2,3$ and $\tau^a$ are the Pauli matrices. The Euler-Lagrange equations of action~\eqref{Action YMH}, with fundamental (real) fields $\Phi^a$ and $A^a_\mu$, are written, respectively, 
\begin{subequations}\label{EL YMH}
 \begin{eqnarray}
 D_\mu D^\mu\Phi&=&-2{\partial V\over\partial|\Phi|}\Phi,\\
 D_\nu F^{\mu\nu}&=&-ie[\Phi, D^\mu\Phi].
\end{eqnarray}
\end{subequations}
Rewriting the Lagrangian density, $\mathcal{L}$, in terms of electric field strength, $E_i=F_{0i}$, and magnetic field strength, $B_i={1\over 2}\epsilon_{ijk}F_{jk}$ with $i=1,2,3$, the effective Lagrangian density is given by
 \begin{equation}
 \mathcal{L}_{eff}\equiv\sqrt{-g}\mathcal{L}=\Tr\left(E_i\right)^2-\Tr\left(B_i\right)^2+\Tr\left(D_0\Phi\right)^2-\Tr\left(D_i\Phi\right)^2-V,
\end{equation}
where the scalar potential $V\geq0$ is still arbitrary. Now consider a BPS Lagrangian density as follows
\begin{equation}\label{Lbps gen YMH}
 \mathcal{L}_{BPS}=2\alpha~\Tr\left(E_iD_i\Phi\right)-2\beta~\Tr\left(B_iD_i\Phi\right)-\gamma~\Tr\left(D_i\Phi\right)^2,
\end{equation}
where $\alpha,\beta,$ and $\gamma$ are arbitrary constants. From both Lagrangian densities, we may obtain
\begin{eqnarray}
 \mathcal{L}_{eff}-\mathcal{L}_{BPS}&=& \Tr\left(E_i-\alpha~ D_i\Phi\right)^2 - \Tr\left(B_i-\beta~ D_i\Phi\right)^2+\Tr\left(D_0\Phi\right)^2 -\left(1-\gamma +\alpha ^2-\beta ^2\right)\Tr\left(D_i\Phi \right)^2-V.
\end{eqnarray}
This form can be obtained by considering $\mathcal{L}_{eff}-\mathcal{L}_{BPS}$ as quadratic equation and completing the square in $E_i,B_i,D_0\Phi,$ and $D_i\Phi$ subsequently.
In the BPS limit, $\mathcal{L}_{eff}-\mathcal{L}_{BPS}=0$, first three terms on the right hand side give us the Bogomol'nyi equations~\cite{Bogomolny:1975de,Manton:2004tk,Weinberg:2012pjx},
\begin{equation}
 E_i=\alpha~ D_i\Phi,\qquad B_i=\beta~ D_i\Phi,\qquad D_0\Phi=0.
\end{equation}
For the fourth term, we can not take $D_i\Phi=0$ otherwise it will lead us to trivial solution and thus we must set $\gamma=1 +\alpha ^2-\beta ^2$ and then $V=0$. As we mentioned previously there are additional constraint equations since the BPS Lagrangian density contains non-boundary terms. The Euler-Lagrange equations of $\mathcal{L}_{BPS}$, respectively, for fields $\Phi$, $A_i$, and $A_0$ are simplied to
 \begin{eqnarray}
 \left(1-\beta^2\right)D_iD_i\Phi&=&0,\\
 \left(1-\alpha^2-\beta^2\right) \left[D_i\Phi,\Phi\right]&=&0,\\
 \alpha~D_iD_i\Phi&=&0.
\end{eqnarray}
There are two possible solutions without additional constraint equations:
\begin{itemize}
 \item BPS monopoles: $\alpha=0, \beta=\pm 1$.
 \item BPS dyons: $\alpha\neq 0\Longrightarrow D_iD_i\Phi=\beta D_iB_i=0$ (by Bianchi Identity) and $\alpha^2+\beta^2=1$.
\end{itemize}
By allowing $\alpha,\beta,$ and $\gamma$ in \eqref{Lbps gen YMH} to be functions of $|\Phi|$, we may also obtain Bogomol'nyi equations for the generalized $SU(2)$ Yang-Mills-Higgs model~\cite{Atmaja:2020iti}.

\subsubsection{Case with ansatz}
Working with PDEs (partial differential equations) are more difficult compared to ODEs (ordinary differential equations). Therefore sometimes it is more practical to turn the PDEs into PDEs with less number of variables, or even if possible to turn them into ODEs, by taking a particular ansatz. Here we will show how to employ the BPS Lagrangian method under ('tHofft-Polyakov) Julia-Zee ansatz~\cite{tHooft:1974kcl,Polyakov:1974ek,Julia:1975ff}, 
\begin{equation}\label{eq:ansatz}
\Phi^a= f(r) {x^a\over r},\qquad A^a_0={j(r)\over e} {x^a\over r},\qquad A^a_i={1-a(r)\over e} \epsilon^{aij} {x^j\over r^2},
\end{equation}
where $x^a\equiv(x,y,z)$, as well as $x^{i,j}\equiv(x,y,z)$, denotes the Cartesian coordinates and $\epsilon^{aij}$ is the Levi-Civita symbol. Under this ansatz the Euler-Lagrange equations \eqref{EL YMH} are effectively written as
\begin{subequations}\label{eq:EoMstd}
\begin{align}
-{1\over r^2}(r^2 f')'+{2 a^2f\over r^2}=&-{dV(f)\over df},\\
-{(r^2 j')'\over er^2}+{2a^2j\over er^2}=&0,\\
{a(a^2-1)\over r^2}+a(e^2f^2-j^2)-a''=&0,
\end{align}
\end{subequations}
where we have defined a differential operator $'\equiv{d\over dr}$.
Now we need to find the effective Lagrangian density and identify its effective fields. It turns out the effective Lagrangian density is proportional to $\sqrt{-g}\mathcal{L}$ under the ansatz \eqref{eq:ansatz}, in spherical coordinates,
\begin{equation}\label{gen eff L}
 \mathcal{L}_{eff}=-r^2\left(\frac{f'^2}{2}+\frac{a^2 f^2}{r^2}\right)+{r^2\over e^2} \left(\frac{j'^2}{2}+\frac{a^2 j^2}{r^2}\right)-{r^2\over e^2} \left(\frac{a'^2}{r^2}+\frac{\left(a^2-1\right)^2}{2 r^4}\right)-r^2V(f),
\end{equation}
and its effective fields are $f, a,$ and $j$ such that Euler-Lagrange equations of $\mathcal{L}_{eff}\left(r,f(r),a(r),j(r),{df(r)\over dr},{da(r)\over dr},{dj(r)\over dr}\right)$ for the effective fields  $f, a,$ and $j$ are given in \eqref{eq:EoMstd} respectively. For this case we consider the effective BPS Lagrangian density with the following form
\begin{equation}
 \mathcal{L}_{BPS}=- {\partial Q\over\partial f} f'-  {\partial Q\over\partial a} a'-  {\partial Q\over\partial j} j',\label{L_BPS eff-1}
\end{equation}
where $Q\equiv Q(f,a,j)$ is an auxiliary function of only the effective fields. We then rewrite $\mathcal{L}_{eff}-\mathcal{L}_{BPS}$ by completing the square in $f',a',$ and $j'$,
\begin{eqnarray}
 \mathcal{L}_{eff}-\mathcal{L}_{BPS}&=&-{r^2\over 2}\left(f'-{Q_f \over r^2}\right)^2-{1\over e^2}\left(a'-{e^2\over 2}Q_a \right)^2+{r^2\over e^2}\left(j'+e^2{Q_j \over r^2}\right)^2\nonumber\\
 &&\frac{e^2 Q_f^2 -e^4 Q_j^2 -\left(a^2-1\right)^2} {2r^2e^2 }+\frac{4 a^2 \left(j^2 -e^2 f^2 \right)+e^4 Q_a^2}{4e^2 }-r^2V,
\end{eqnarray}
with $Q_f\equiv{\partial Q\over\partial f}, Q_a\equiv{\partial Q\over\partial a},$ and $Q_j\equiv{\partial Q\over\partial j}$.
In the BSP limit, $\mathcal{L}_{eff}-\mathcal{L}_{BPS}=0$ implies Bogomol'nyi equations
\begin{equation}
  f'={Q_f\over r^2},\qquad a'={e^2\over 2} Q_a,\qquad j'=-e^2 {Q_j \over r^2}.
\end{equation}
and 
\begin{equation}
\frac{e^2 Q_f^2 -e^4 Q_j^2 -\left(a^2-1\right)^2} {2r^4e^2 }+\frac{4 a^2 \left(j^2 -e^2 f^2 \right)+e^4 Q_a^2}{4r^2 e^2 }-V=0
\end{equation}
which can be solved by expanding the left hand side in explicit radial coordinate $r$ and then setting all the ``coefficients'' to be zero, $V=0$,
\begin{equation}
 V=0,\qquad Q_a=\pm {2\over e^2} a \sqrt{e^2f^2-j^2},\qquad e^2 Q_f^2-e^4 Q_j^2=\left(a^2-1\right)^2.
\end{equation}
Therefore we can fix the auxiliary function $Q=\pm {1\over e^2}(a^2-1)\sqrt{e^2f^2-j^2}$ and the Bogomol'nyi equations now become~\cite{Bogomolny:1975de,Manton:2004tk,Weinberg:2012pjx}
\begin{equation}
 f'=\pm{(a^2-1)f\over r^2\sqrt{e^2f^2-j^2}},\qquad a'=\pm a \sqrt{e^2f^2-j^2},\qquad j'=\pm {(a^2-1)j \over r^2\sqrt{e^2f^2-j^2}}.
\end{equation}
There are no additional constraint equations since the Euler-Lagrange equations of $\mathcal{L}_{BPS}$ \eqref{L_BPS eff-1} are trivially satisfied. Furthermore from the Bogomol'nyi equations we may have an equation ${j'/ f'}={j/f}$ whose solution is $j(r)=\sigma f(r)$, with $\sigma$ is an integration constant.

\section{Four-dimensional Gravity}
The four-dimensional Einstein-Hilbert action
\begin{equation}
 S_{EH}={1\over 2\kappa}\int d^4x\sqrt{-g} \left(R-2\Lambda\right),
\end{equation}
where $\Lambda$ is the cosmological constant and $\kappa=8\pi G$ is the Einstein gravitational constant.
In general, it requires additional Gibbons-Hawking-York boundary terms as such the resulting Euler-Lagrange equations are the Einstein equations~\cite{York:1972sj,Gibbons:1976ue}. The total action is given by
\begin{align}\label{EH action}
 S=&S_{EH}+S_{GHY}\nonumber\\
  =&{1\over 2\kappa}\int_\mathcal{M} d^4x\sqrt{-g}\left(R-2\Lambda\right)+{1\over \kappa}\int_{\partial\mathcal{M}}d^3y\epsilon\sqrt{h}K-{1\over \kappa}\int_{\partial\mathcal{M}}d^3y\epsilon\sqrt{h}K_0,
\end{align}
where $h_{ab}$ is the induced metric; $K$ is the trace of the extrinsic curvature; and $\epsilon=+1$ if the normal of boundary manifold $\partial\mathcal{M}$ is spacelike and  $\epsilon=-1$ if the normal of boundary manifold $\partial\mathcal{M}$ is timelike, with $y^a$ are the coordinates on the boundary manifold $\partial\mathcal{M}$. Here the last term is added to remove singularity part of the Gibbons-Hawking-York terms. 

\subsection{Static spherically symmetric}
In this case the ansatz for the four-dimensional metric is 
\begin{equation}\label{ansatz}
 ds^2=-A(r)dt^2+B(r)dr^2+C(r)d\Omega_2^2,
\end{equation}
where $A,B,C\geq0$. Non-trivial parts of the Einstein equations, $R_{\mu\nu}-{1\over 2}g_{\mu\nu}R+\Lambda g_{\mu\nu}=0$, in this ansatz are simplified to
\begin{subequations}\label{Einstein}
 \begin{eqnarray}
 2 C(r) A'(r) C'(r)+A(r) \left(C'(r)^2-4 B(r) C(r)\left(1-\Lambda~ C(r)\right)\right)=0,\\
 2 C(r) B'(r) C'(r)+B(r) \left(C'(r)^2-4 C(r) C''(r)\right)+4 B(r)^2 C(r)\left(1-\Lambda~ C(r)\right)=0,\\
 B(r) \left(-A(r) C(r) \left(A'(r) C'(r)+2 A(r) C''(r)\right)+C(r)^2 \left(A'(r)^2-2 A(r) A''(r)\right)+A(r)^2 C'(r)^2\right)\nonumber\\
 +A(r) C(r) B'(r) \left(C(r) A'(r)+A(r) C'(r)\right)-4\Lambda~ A(r)^2 B(r)^2 C(r)^2=0.
\end{eqnarray}
\end{subequations}
Now we are going to find the Bogomol'nyi-like equations, or first-order derivative equations, for functions $A,B,$ and $C$ using the BPS Lagrangian method~\cite{Atmaja:2015umo}. The BPS Lagrangian method requires a Lagrangian density, which shall be called effective Lagrangian density, that give us the Einstein equations \eqref{Einstein}. To do so, we use the Gaussian normal coordinates and pick a boundary manifold as timelike hypersurface, with spacelike normal vector or $\epsilon=1$, at spatial infinity assumed to be at $r\to\infty$. The reduced Einsten-Hilbert action is given by
\begin{equation}\label{E-H reduced}
 S_{RedEH}\propto\int dr \frac{2 C(r) A'(r) C'(r)+A(r) \left(4 B(r) C(r)\left(1-\Lambda~ C(r)\right)+C'(r)^2\right)}{\sqrt{A(r) B(r) C(r)^2}}=\int dr \mathcal{L}_{eff}.
\end{equation}
One can check that the Euler-Lagrange equations of this reduced action are indeed the Einstein equations \eqref{Einstein}. Derivation of Euler-Lagrange equations \eqref{Einstein} from the reduced action \eqref{E-H reduced} is rather naive by assuming the radial coordinate $r$ to act as time coordinate and the variation is taken over the effective fields $(A,B,C)$. We will also use these assumptions when applying the BPS Lagrangian method further.
Following prescriptions of the BPS Lagrangian method, we pick a standard BPS Lagrangian density which contains linear terms in $A'(r),B'(r),$ and $C'(r)$ as follows~\footnote{This form of BPS Lagrangian density will contain only the ``boundary'' terms if $X_0=0, X_a={\partial Q\over\partial A}, X_b={\partial Q\over\partial B},$ and $X_c={\partial Q\over\partial C}$.}
\begin{equation}\label{Lbps}
 \mathcal{L}_{BPS}=X_0(A,B,C)+X_a(A,B,C) A'(r)+ X_b(A,B,C) B'(r)+ X_c(A,B,C) C'(r),
\end{equation}
where $X_0,X_a,X_b,$ and $X_c$ are auxilliary functions of $A,B,$ and $C$. Our task now is to find explicit form of these functions.
We then rewrite $\mathcal{L}_{eff}-\mathcal{L}_{BPS}$ as follows
\begin{eqnarray}
&&-{C\over A\sqrt{AB}} \left(A'(r)-\frac{\sqrt{A B}}{2 C}\left(X_cC-X_aA\right)\right)^2+\sqrt{A\over B}\left({C'(r)\over C}+{1\over A}\left(A'(r)-\frac{X_c}{2}\sqrt{A B}\right)\right)^2\nonumber\\
&&-B'(r) X_b-X_0+{\sqrt{AB}\over 4C}\left(X_a^2A-2(X_a X_c-8\left(1-\Lambda~ C\right))C\right).
\end{eqnarray}
In the BPS limit, where $\mathcal{L}_{eff}-\mathcal{L}_{BPS}=0$, we can extract Bogomol'nyi-like equations for $A$ and $C$, respectively,
\begin{subequations}\label{BE A and C}
 \begin{eqnarray}
  A'(r)&=&\frac{\sqrt{A B}}{2 C}\left(X_cC-X_aA\right),\\
  C'(r)&=&-{C\over A}\left(A'(r)-\frac{X_c}{2}\sqrt{A B}\right)
 \end{eqnarray}
\end{subequations}
and the remaining terms is
\begin{equation}
 {\sqrt{AB}\over 4C}\left(X_a^2A-2(X_a X_c-8\left(1-\Lambda~ C\right))C\right)-B'(r) X_b=0.
\end{equation}
Ones can consider this equation as first-order differential equation of $B$ which could be later identified as Bogomol'nyi-like equation for $B$. This however contradicts with our previous results that there are only Bogomol'nyi-like equations for $A$ and $C$. In order to avoid this contradiction we must set $X_b=0$ which later implies
\begin{equation}
 X_c=\frac{A B \left(X_a^2 A +16 C\left(1-\Lambda~ C\right)\right)-4X_0C \sqrt{A B} }{2X_a A B C }.
\end{equation}
Next, we must also consider additional constraint equations that are Euler-Lagrange equations of $S=\int dr \mathcal{L}_{BPS}$. The constraint equation for $B$ yields $X_0=0$. The remaining constraint equations can be simplified to
\begin{equation}\label{CE}
 {4 C\over \sqrt{A B}} {\partial X_a\over \partial B} X_a~ B'(r)=X_a^3-2C {\partial X_a\over\partial C}X_a^2-\left(16 C\left(1-\Lambda~ C\right)-X_a^2 A \right) {\partial X_a\over\partial A}.
\end{equation}
This simplified constraint equation is a first-order differential equation of $B$. Using explicit functions of $X_0,X_b,$ and $X_c$ the Bogomol'nyi-like equations \eqref{BE A and C} can be simplified to
\begin{subequations}\label{BE A and C new}
 \begin{eqnarray}
  A'(r)&=&\sqrt{A B} \left(\frac{4\left(1-\Lambda~ C\right)}{X_a(A,B,C)}-\frac{A}{4 C}X_a(A,B,C)\right),\label{BE A}\\
  C'(r)&=&\frac{\sqrt{A B}}{2} X_a(A,B,C),\label{BE C}
 \end{eqnarray}
\end{subequations}
which, together with the constraint equation \eqref{CE}, satisfy the Einstein equations \eqref{Einstein}. However those equations are impractical when we want to find their explicit solutions because there is still one auxilliary function $X_a$ needs to be determined.

\subsubsection{Schwarzschild black hole: $C=r^2$}
In this case the solution for $X_a$, from the Bogomol'nyi-like equation \eqref{BE C}, is given by $X_a={4r\over\sqrt{AB}}$. With this explicit function of $X_a$, the Bogomol'nyi-like equation for $A$ \eqref{BE A} and the contraint equation \eqref{CE} are now
\begin{subequations}
 \begin{eqnarray}
  A'(r)&=&{A}{B\left(1-\Lambda~ r^2\right)-1\over 2r},\\
   B'(r)&=&-{B}{B\left(1-\Lambda~ r^2\right)-1\over 2r}.
 \end{eqnarray}
\end{subequations}
From the ratio $A'(r)/B'(r)$, we may conclude the functions $A$ and $B$ are related by $A={c_a\over B}$, with $c_a$ is a real constant. Solutions to $A$ and $B$ are
\begin{eqnarray}
 A&=&c_a{(r-{r^3\over 3}\Lambda+c_b)\over r },\\
 B&=&{r\over r-{r^3\over 3}\Lambda+ c_b},
\end{eqnarray}
which are the Schwarzschild solutions. In particular the Schwarzschild black hole is obtained by setting $\Lambda=0, c_a=1,$ and $c_b=-2 M$, where $M$ is the mass of gravitational source.

\subsubsection{General solutions}\label{uniqueness solution 1}
Using the Bogomol'nyi-like equations \eqref{BE A and C new}, we can recast the constraint equation \eqref{CE} to be
\begin{equation}
 {dX_a\over dr}={\sqrt{AB}\over 4C}X_a^2.
\end{equation}
Comparing it with the Bogomol'nyi-like equation for $C$ \eqref{BE C}, we obtain $X_a=cx_a\sqrt{C}$, with $cx_a$ is a real constant. Furthermore using explicit solution of $X_a$ and comparing both Bogomol'nyi-like equations \eqref{BE A and C new}, we may obtain a differential equation
\begin{equation}
 2C{dA\over dC}={16\over cx_a^2}\left(1-\Lambda C\right)-A
\end{equation}
whose solution is 
\begin{equation}
 A=\frac{16}{3 cx_a^2}\left(3-\Lambda~ C\right)+\frac{c_a}{\sqrt{C}},
\end{equation}
with $c_a$ is a real constant. In order to find solution for $B$, we can not use the constraint equation \eqref{CE} since it becomes trivial when we substitute the explicit solution of $X_a$. We can use the Bogomol'nyi's equation \eqref{BE C} to obtain solution for $B$ as follows
\begin{equation}\label{sol B 1}
 B={4\over cx_a^2}{1 \over A~ C}C'(r)^2.
\end{equation}
So the solution for $B$ depends on explicit function of $C(r)$. Here we can see $B\propto A^{-1}$.

\subsection{Static Electrovac}\label{uniqueness solution 2}
Now consider incorporating electromagnetic field into the Hilbert-Einstein action such that the total action is
\begin{equation}\label{electrovac action}
 S=\int d^4x \sqrt{-g}\left({1\over 2\kappa}\left(R-2\Lambda\right)-{1\over 4}F_{\mu\nu}F^{\mu\nu}\right),
\end{equation}
with $F_{\mu\nu}=D_\mu A_\nu-D_\nu A_\mu$ and $D_\mu$ is the covariant derivative. For our convinient we set $\kappa=1$ from now on. A general static spherical symmetric ansatz for electromagnetic field is given by
\begin{equation}\label{ansatz-2}
 A_\mu=\left(A_t(r),A_r(r), A_\theta \sin\theta, A_\phi \cos\theta\right),
\end{equation}
with $A_\theta$ and $A_\phi$ are real constants. Within this ansatz the function $A_r(r)$ and constant $A_\theta$ do not appear anywhere in the action so we may safetly set them to zero, $A_r(r)=A_\theta=0$. Under the ansatzes \eqref{ansatz} and \eqref{ansatz-2}, the reduced action are given by
\begin{equation}
 S_{Red}=\int dr\mathcal{L}_{eff}=\int dr \frac{2 C(r) \left(A'(r) C'(r)+C(r) A_t'(r)^2\right)+A(r) \left(C'(r)^2-2 B(r) \left(A_\phi^2+2 C(r) (\Lambda  C(r)-1)\right)\right)}{\sqrt{A(r) B(r)}C(r)}.
\end{equation}
Again one can check that Euler-Lagrange equations of this effetive Lagrangian $\mathcal{L}_{eff}$ satisfy the Einstein equations and the Maxwell equations, $D^\mu F_{\mu\nu}=0$.
The BPS Lagrangian density is taken to be
\begin{equation}\label{Lbps-1}
 \mathcal{L}_{BPS}=X_0(A,B,C,A_t)+X_a(A,B,C,A_t) A'(r)+ X_b(A,B,C,A_t) B'(r)+ X_c(A,B,C,A_t) C'(r)+ X_t(A,B,C,A_t) A_t'(r),
\end{equation}
where $X_0,X_a,X_b,X_c,$ and $X_t$ are auxilliary functions of $A(r),B(r),C(r),$ and $A_t(r)$.
The equation $\mathcal{L}_{eff}-\mathcal{L}_{BPS}=0$ can be rewritten to be
\begin{eqnarray}\label{BPS limit}
 &&\frac{2 C}{\sqrt{AB}}\left(A_t'(r)-\frac{\sqrt{A B} X_t}{4 C}\right)^2+\frac{1}{C}\sqrt{A\over B}\left(C'(r)+\frac{1}{2} C \sqrt{\frac{B}{A}} \left(\frac{2 A'(r)}{\sqrt{A} \sqrt{B}}-X_c\right)\right)^2-\frac{C}{A^{3/2} \sqrt{B}}\left(A'(r)-\frac{\left(A^{3/2} \sqrt{B}\right)}{2 C}\left(\frac{C X_c}{A}-X_a\right)\right)^2\nonumber\\
 &-&{\sqrt{AB}\over 8 C} \left(4 C (X_a X_c+8 \Lambda  C-8)+X_t^2+16 A_\phi^2\right)-X_0+{A^{3/2} \sqrt{B}\over 4 C} X_a^2-B'(r) X_b=0.
\end{eqnarray}
In the BPS limit, the first line of equation \eqref{BPS limit} will give us Bogomol'nyi-like equations for $A_t, C,$ and $A$ while the second line must zero such that, for the same reason as in the previous case, $X_b=0$ and thus 
\begin{equation}
 X_c=-\frac{\sqrt{AB} \left(-2 A~ X_a^2+X_t^2+16 \left(A_\phi^2+2 C (C~ \Lambda -1)\right)\right)+8 C~ X_0}{4 \sqrt{A} \sqrt{B} C~X_a}.
\end{equation}
Substituting $X_b$ and $X_c$, the Bogomol'nyi-like equations for $A_t, C,$ and $A$ now become
\begin{subequations}\label{BE Electrovac}
 \begin{eqnarray}
  A_t'(r)&=&\frac{\sqrt{A~ B}}{4 C} X_t,\label{BEAt}\\
  C'(r)&=&\frac{1}{2} \sqrt{A~B}~ X_a,\label{BEC}\\
  A'(r)&=&-\frac{\sqrt{A~B} \left(2 A~ X_a^2+X_t^2+16 \left(A_\phi^2+2 C (C \Lambda -1)\right)\right)}{8 C X_a}+ {X_0\over X_a}.\label{BEA}
 \end{eqnarray}
\end{subequations}
The remaining unknown functions $X_a$ and $X_t$ can be determined from the Euler-Lagrange equations of BPS Lagrangian density \eqref{Lbps-1}. The constraint equation for $B$ yields $X_0=0$ which further simplifies the constraint equation for $A_t$ to be ${dX_t\over dr}=0$, or $X_t=cx_t$ is a real constant. The remaining constraint equations, for $A$ and $C$, can be simplified to
\begin{equation}
 8 C~ X_a {\partial X_a \over \partial B}B'(r)=\sqrt{A~B} \left({\partial X_a \over \partial A} \left(2 A~ X_a^2+16 A_\phi^2+32 C (C \Lambda -1)+cx_t^2\right)+2 X_a \left(-cx_t {\partial X_a \over \partial A_t}-2 C {\partial X_a \over \partial C} X_a+X_a^2\right)\right).\label{CEB}
\end{equation}
Using the Bogomol'nyi-like equations \eqref{BE Electrovac}, we can recast this constraint equation to be
\begin{equation}
 {dX_a \over dr}={\sqrt{A~B}\over 4 C}X_a^2={X_a\over 2C} {dC \over dr}
\end{equation}
whose solution is $X_a=cx_a\sqrt{C}$, with $cx_a$ is the integration constant. Using the explicit solutions $X_a$ and $X_t$, and comparing the Bogomol'nyi-like equations \eqref{BEAt} and \eqref{BEC} we then obtain
\begin{equation}
 A_t=c_t-{cx_t \over cx_a}C^{-1/2},
\end{equation}
with $c_t$ is the integration constant. Furthermore comparing the Bogomol'nyi-like equations \eqref{BEC} and \eqref{BEA}, we may obtain a differential equation
\begin{equation}
 -4 cx_a^2 C^2{dA\over dC}=16 \left(A_\phi^2+2 C (\Lambda ~C-1)\right)+2 cx_a^2 A~C+cx_t^2
\end{equation}
whose solution is given by
\begin{equation}
 A=\frac{1}{6 cx_a^2 C}\left(48 A_\phi^2-32 C (\Lambda~C -3)+3 cx_t^2\right)+\frac{c_a}{\sqrt{C}},
\end{equation}
with $c_a$ is a real constant. As in the previous case, the solution for $B$ can be obtained from the Bogomol'nyi equation \eqref{BEC},
\begin{equation}\label{sol B 2}
 B={4\over cx_a^2}{1\over A~C}C'(r)^2.
\end{equation}
If we take $C(r)=r^2$ and set $cx_a=4,cx_t=4 Q,$ and $c_a=-2M$ then we get the Reissner-Nordstr\"{o}m black hole solutions,
\begin{equation}
 ds^2=-\left(1-{r^2\over 3}\Lambda-{2M\over r}+{Q^2+A_\phi^2\over 2 r^2}\right)dt^2+\left(1-{r^2\over 3}\Lambda-{2M\over r}+{Q^2+A_\phi^2\over 2 r^2}\right)^{-1}dr^2+r^2 d\Omega_2^2,
\end{equation}
where $M$ is the black hole mass, $Q$ is the electric charge, and $A_\phi$ is the magnetic charge.

\subsection{Einstein-Scalar Gravity}\label{ES gravity}
An action for four-dimensional Einstein-Scalar gravity is given by
\begin{equation}\label{action ES gravity}
 S=\int~dx^4 \sqrt{-g}\left({1\over 2\kappa}\left(R-2\Lambda\right)-{1\over 2}D_\mu\phi D^\mu\phi-V(\phi)\right),
\end{equation}
with $V$ is a generic scalar potential. We take an ansatz for the real scalar field $\phi\equiv\phi(r)$ such that, together with the ansatz \eqref{ansatz}, the reduced action is given by
\begin{eqnarray}\label{eff Lagrangian ES gravity}
 S_{red}&=&\int dr\mathcal{L}_{eff}=\int dr \left(\frac{2 C~ A'(r) C'(r)+ 4 A B C \left(1- \Lambda  C\right)+A~ C'(r)^2}{\sqrt{A~B}~C}-\sqrt{A B}C \left(\frac{2}{B}\phi'(r)^2+4V(\phi)\right)\right).
\end{eqnarray}
In this case we consider BPS Lagrangian density as follows
\begin{equation}
 \mathcal{L}_{BPS}=X_0(A,B,C,\phi)+X_a(A,B,C,\phi) A'(r)+ X_b(A,B,C,\phi) B'(r)+ X_c(A,B,C,\phi) C'(r)+X_\phi(A,B,C,\phi) \phi'(r),
\end{equation}
where $X_0,X_a,X_b,X_c,$ and $X_\phi$ are auxilliary functions of $A(r),B(r),C(r),$ and $\phi(r)$.
In the BPS limit, $\mathcal{L}_{eff}-\mathcal{L}_{BPS}=0$ can be rewritten as
\begin{eqnarray} \label{BPS limit-2}
-\frac{2\sqrt{A}C}{\sqrt{B}} \left(\phi'(r)+\frac{\sqrt{B}}{4\sqrt{A}C} X_\phi\right)^2-\frac{C}{A^{3/2} B^{1/2}}\left(A'(r)-\frac{A^{3/2} B^{1/2}}{2 C}\left(\frac{C}{A}X_c -X_a\right)\right)^2+\frac{A^{1/2}}{B^{1/2}~C}\left(C'(r)+\frac{B^{1/2}C}{2 A^{1/2}}\left(\frac{2}{\sqrt{A~B}}A'(r)-X_c\right)\right)^2\nonumber\\
 -B'(r) X_b-X_0+\frac{\sqrt{B}}{8 \sqrt{A} C} \left(-4 A C (X_a X_c+8 C (\Lambda +V)-8)+2 A^2 X_a^2+X_\phi^2\right)=0.\nonumber\\
\end{eqnarray}
The first line of equation \eqref{BPS limit-2} will give us Bogomol'nyi-like equations while the second line must be set to zero such that, similar to previous cases, $X_b=0$ and 
\begin{equation}
X_c=\frac{1}{4 A C X_a}\left(2 A^2 X_a^2-\frac{8 \sqrt{A} C}{\sqrt{B}} X_0+X_\phi^2-32 A C^2 V-32 A C ( \Lambda C -1)\right).
\end{equation}
Substituting explicit functions $X_b$ and $X_c$, the Bogomol'nyi-like equations for $\phi, C,$ and $A$ become
\begin{subequations}\label{BE Electrovac-2}
 \begin{eqnarray}
  \phi'(r)&=&-{1\over 4 C}\sqrt{B\over A} X_\phi,\label{BEph-2}\\
  C'(r)&=&\frac{\sqrt{A~ B}}{2} X_a,\label{BEC-2}\\
  A'(r)&=&\frac{-2 A^2 \sqrt{B}X_a^2-8 \sqrt{A} C X_0+\sqrt{B}X_\phi^2-32 A \sqrt{B} C (C (\Lambda +V)-1)}{8 \sqrt{A} C X_a}.\label{BEA-2}
 \end{eqnarray}
\end{subequations}
The remaining unknown functions $X_0, X_a,$ and $X_\phi$ can be determined from Euler-Lagrange equations of the BPS Lagrangian density $\mathcal{L}_{eff}$. The constraint equation for $B$ yields $X_0=0$. Furthermore using this explicit $X_0$, the constraint equations for $A$ and $\phi$ can be simplified to, respectively,
\begin{subequations}
 \begin{eqnarray}
 {X_a}'(r)&=&-\frac{\sqrt{B}}{8A^{3/2}C}\left(X_\phi^2-2 A^2 X_a^2\right),\label{CE Xa}\\
 {X_\phi}'(r)&=&-4\sqrt{AB} C~ V'(\phi),\label{CE Xphi}
\end{eqnarray}
\end{subequations}
These equations further implies the constraint equation for $C$ to be trivial.

Let us consider solution for which $C=r^2$. In this case solutions are obtained by setting $X_a=4\sqrt{C\over A~B}$ which further simplifies the Bogomol'nyi-like equation for $A$ and , from the constraint equation \eqref{CE Xa}, implies a constraint equation for $B$, respectively
\begin{eqnarray}
 A'(r)&=&\frac{1}{32 C^{3/2}}\left(B X_\phi^2-32 A C (B (C \Lambda -1)+B C V+1)\right),\\
 B'(r)&=&\frac{B}{32 A C^{3/2}}\left(B X_\phi^2+32 A C (B (C \Lambda -1)+B C V+1)\right).
\end{eqnarray}
Using these equations we can write 
\begin{equation}
 {d\log(AB) \over dr}=\sqrt{C}\phi'(r)^2.
\end{equation}
Now take an example where solution for the scalar field is a Coulomb-like solution where $\phi={Q_s\over r}$ with $Q_s$ is a real constant.
This implies $X_\phi=4\phi  \sqrt{A~C\over B}$ and
\begin{equation}
 {d\log(AB) \over dr}={\phi^2\over\sqrt{C}}=Q_s^2 {1\over r^3}
\end{equation}
whose solution is
\begin{equation}
 A=c_a{e^{-{\phi^2\over 2}}\over B},
\end{equation}
with $c_a$ is an integration constant which we will set to unity, $c_a=1$. Since all the auxiliary functions have been fixed, the remaining equations to be solved, in terms of $\phi$, are
\begin{eqnarray}\label{Coulomb scalar}
B'(\phi)&=&-\frac{B}{2 \phi ^3} \left(2Q_s^2 B(\Lambda+V)-2 (B-1) \phi ^2+\phi ^4\right),\nonumber\\
 V'(\phi)&=&\phi \left(-\frac{(B-1) \phi ^2}{Q_s^2 B}+\Lambda +V\right).
\end{eqnarray}
Solutions to these equations are
\begin{eqnarray}\label{New BH1}
 B^{-1}&=&{1\over \phi^2}\left(e^{\frac{\phi ^2}{4}} \phi  \left(-2 \sqrt{\pi } \text{erf}\left(\frac{\phi }{2}\right)+c_2 Q_s^2\right)+e^{\frac{\phi ^2}{2}} \left(-\sqrt{\pi } c_2 Q_s^2 \text{erf}\left(\frac{\phi }{2}\right)+\pi  \text{erf}\left(\frac{\phi }{2}\right)^2+2 c_1 Q_s^2\right)-4\right),\\
 V&=&-\Lambda+{\phi^2+12\over Q_s^2}-{3\over Q_s^2} e^{\frac{\phi ^2}{4}} \phi  \left(-2 \sqrt{\pi } \text{erf}\left(\frac{\phi }{2}\right)+c_2 Q_s^2\right)+{\left(\phi ^2-6\right)\over 2 Q_s^2}e^{\frac{\phi ^2}{2}}  \left(-\sqrt{\pi } c_2 Q_s^2 \text{erf}\left(\frac{\phi }{2}\right)+\pi  \text{erf}\left(\frac{\phi }{2}\right)^2+2 c_1 Q_s^2\right), \nonumber\\
\end{eqnarray}
where $\text{erf}(z)={2\over\sqrt{\pi}}\int_0^z e^{-t^2}dt$ is the Error function, with $c_1$ and $c_2$ are constant. At $r\to\infty$, we have 
\begin{eqnarray}
 A(r\to\infty)&\sim&-\lambda r^2 +1-{Q_s^3 c_2\over 6 r}+O\left(\left( 1\over r\right)^3\right),\\
 V(r\to\infty)&\sim& 3 \lambda -\Lambda +\frac{\lambda  Q_s^2}{r^2}+\frac{\lambda  Q_s^4}{8 r^4}-\frac{c_2 Q_s^5}{30 r^5}+ O\left(\left( 1\over r\right)^6\right),
\end{eqnarray}
with $\lambda=\frac{4}{Q_s^2}-2 c_1$ is an effective cosmological constant. The existance of asymptotically flat black holes, with $\lambda=0$, depend on the value of $c_2$.
\begin{figure}[H]
    \centering
    \includegraphics[width=0.8\textwidth]{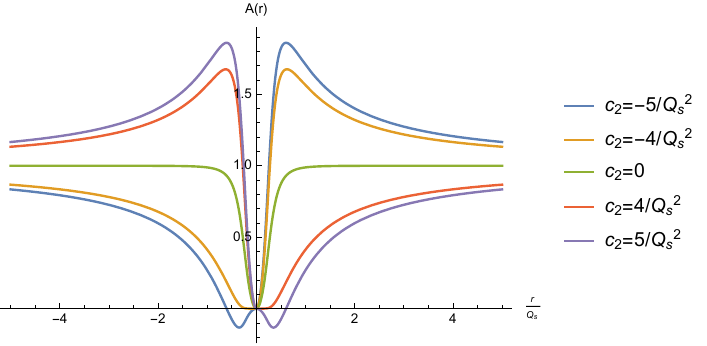}
    \caption{All curves in the region $r/Q_s<0$ correspond to $Q_s<0$. The horizon is larger as we increase (decrease) the value of $c_2$ for fixed $Q_s>0$ ($Q_s<0$). The horizon radius for $c_2=|4|/Q_s^2$ is $r_h\approx 1.03 x 10^{-8} |Q_s|$ and it increases by an order of $10^7$ for $c_2=|5|/Q_s^2$, which is $r_h\approx 0.6~|Q_s|$.}
    \label{fig:A}
\end{figure}

\begin{figure}[H]
    \centering
    \includegraphics[width=0.4\textwidth]{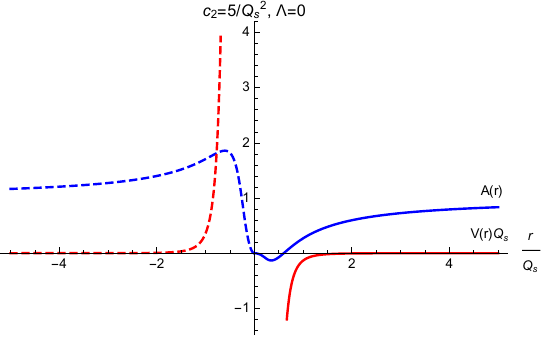}\qquad\qquad
    \includegraphics[width=0.4\textwidth]{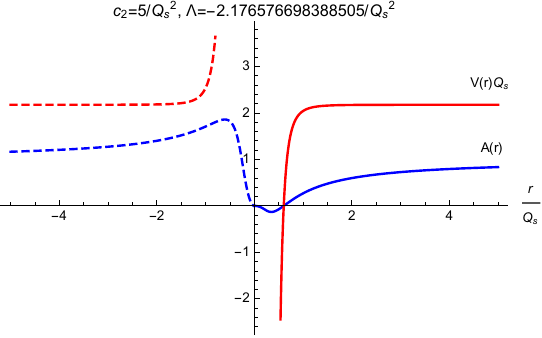}
    \caption{In the left figure, with $\Lambda=0$ and $Q_s>0$, the potential is always negative in the whole space even though the space is cut-off from below at $r_h\approx 0.6~Q_s$ in the presence of horizon. However it can be lift-up such that it can have positive value for all $r>r_h$ when the cosmological constant is negative, $\Lambda\lesssim -2.2/Q_s^2$, as shown in the right figure.}
    \label{fig:Lambda}
\end{figure}

\section{Einstein-Maxwell-Scalar gravity in $n-$dimensions}
The action for Einstein-Maxwell-Scalar gravity is taken to be
\begin{equation}
 S=\int~dx^n \sqrt{-g}\left({1\over 2\kappa}\left(R-2\Lambda\right)-{1\over 4}F_{\mu\nu}F^{\mu\nu}-{1\over 2}D_\mu\phi D^\mu\phi-V(\phi)\right).
\end{equation}
the $n$-dimensional Einstein's equations, Maxwell's equations, and Scalar's equation are respectively given  by
\begin{subequations}\label{Eqns E-M-S}
\begin{eqnarray}
  R_{\mu\nu}-{1\over 2}g_{\mu\nu}R+\Lambda g_{\mu\nu}&=&\kappa\left({F_\mu}^\gamma F_{\nu\gamma}-{g_{\mu\nu}\over 4}F_{\gamma\sigma}F^{\gamma\sigma}+D_\mu\phi D_\nu\phi-{g_{\mu\nu}\over 2}D_\gamma\phi D^\gamma\phi-g_{\mu\nu} V(\phi)\right),\\
 D^\mu F_{\mu\nu}&=&0,\\
 D^\mu D_\mu \phi&=&V'(\phi),
 \end{eqnarray}
\end{subequations}
with $D_\mu$ is the covariant derivative.
We take ansatz for the $n$-dimensional metric to be
\begin{equation}\label{ansatz-1}
 ds^2=-A(r)dt^2+B(r)dr^2+C(r)d\Omega_{n-2}^2,
\end{equation}
where $A,B,C\geq0$ and $n>4$. For the metric of unit hyper-sphere, we follow convention in \cite{Tangherlini:1963bw} as below:
\begin{equation}
 d\Omega_{n-2}^2=\sum_{i=1}^{n-2}\left(\prod_{l=1}^{i-1}\sin^2\theta_l~ d\theta_i^2\right).
\end{equation}
Here we take ansatz for the $n$-dimensional gauge field as follows~\cite{Tangherlini:1963bw,Kodama:2003kk}
\begin{equation}\label{ansatz-3}
 A_\mu=\left(A_t(r),0,\dots,0\right)
\end{equation}
and for the scalar field to be static and spherically symmetric, $\phi\equiv\phi(r)$.
Using those ansatzes, the equations \eqref{Eqns E-M-S} can be obtained from the following reduced action
\begin{eqnarray}
 S_{red}&=&\int dr\mathcal{L}_{eff}\\
 &=&\int dr \left(\frac{2 (n-2) C~ A'(r) C'(r)+ 4 A B C \left((n-3)(n-2)-2 \Lambda  C\right)+(n-3) (n-2) A~ C'(r)^2}{8\kappa \sqrt{A~B}~C^{3-\frac{n}{2}} }\right.\\ \nonumber
 &&\left.\qquad+\sqrt{A B C^{n-2}} \left(\frac{A_t'(r)^2}{2 A B}-\frac{\phi'(r)^2}{2 B}-V(\phi)\right)\right).
\end{eqnarray}
Here again without loos of generality we may set $\kappa=1$. We consider the BPS Lagrangian density to be
\begin{eqnarray}
 \mathcal{L}_{BPS}&=&X_0(A,B,C,A_t,\phi)+X_a(A,B,C,A_t,\phi) A'(r)+ X_b(A,B,C,A_t,\phi) B'(r)+ X_c(A,B,C,A_t,\phi) C'(r)\nonumber\\
 &&+X_t(A,B,C,A_t,\phi) A_t'(r)+X_\phi(A,B,C,A_t,\phi) \phi'(r),
\end{eqnarray}
where $X_0,X_a,X_b,X_c,X_t,$ and $X_\phi$ are auxilliary functions of $A(r),B(r),C(r),At(r),$ and $\phi(r)$.
In the BPS limit, $\mathcal{L}_{eff}-\mathcal{L}_{BPS}=0$ can be rewritten as
\begin{eqnarray} \label{BPS limit-1}
\frac{C^{\frac{n}{2}-1}}{2\sqrt{A~B}} \left(A_t'(r)-\frac{\sqrt{A~B}}{C^{\frac{n}{2}-1}} X_t\right)^2-\frac{\sqrt{A}C^{\frac{n}{2}-1}}{2\sqrt{B}} \left(\phi'(r)+\frac{\sqrt{B}}{\sqrt{A}C^{\frac{n}{2}-1}} X_\phi\right)^2
 \nonumber\\
 -\frac{(n-2) C^{\frac{n}{2}-1}}{8 (n-3) A^{3/2} B^{1/2}}\left(A'(r)-\frac{4(n-3) A^{3/2} B^{1/2}}{(n-2) C^{\frac{n}{2}-1}}\left(\frac{C}{(n-3) A}X_c -X_a\right)\right)^2\nonumber\\
 +\frac{(n-3) (n-2) A^{1/2}}{8 B^{1/2}~C^{3-\frac{n}{2}}}\left(C'(r)+\frac{4B^{1/2}C^{3-\frac{n}{2}}}{(n-3) (n-2) A^{1/2}}\left(\frac{(n-2) C^{\frac{n}{2}-2}}{4\sqrt{A~B}}A'(r)-X_c\right)\right)^2-B'(r) X_b-X_0\nonumber\\
 +\frac{\sqrt{B}~C^{-\frac{n}{2}-2}}{2 (n-2)\sqrt{A}}\left(C^3 \left(A \left(4 (n-3) A X_a^2-(n-2) X_t^2-8 C X_a X_c \right)+(n-2) X_\phi^2\right)+(n-2) A C^n \left(-2 C (\Lambda +V)+(n-3)(n-2)\right)\right)=0.\nonumber\\
\end{eqnarray}
The first three terms in left hand side of equation \eqref{BPS limit-1} will give us Bogomol'nyi-like equations while the remaining terms must be set to zero such that $X_b=0$ and 
\begin{equation}
X_c=\frac{\sqrt{B} \left(4 A^2 C^3 (n-3) X_a^2+(n-2) \left(C^3 \left(X_\phi^2-A X_t^2\right)+A C^n ((n-3)(n-2)-2 C \Lambda)\right)\right)-2 \sqrt{A} (n-2) C^{\frac{n}{2}+2}\left(X_0+\sqrt{AB} C^{\frac{n}{2}+2} V\right)}{8 A \sqrt{B} C^4 X_a}
\end{equation}

Substituting $X_b$ and $X_c$, the Bogomol'nyi-like equations for $A_t,\phi,C,$ and $A$ respectively become
\begin{subequations}\label{BE Electrovac-1}
 \begin{eqnarray}
  A_t'(r)&=&\sqrt{A~ B}~C^{1-{n\over 2}} X_t,\label{BEAt-1}\\
  \phi'(r)&=&-\sqrt{B\over A}C^{1-\frac{n}{2}} X_\phi,\label{BEph-1}\\
  C'(r)&=&\frac{4\sqrt{A~ B}}{(n-2)}C^{2-{n\over 2}} X_a,\label{BEC-1}\\
  A'(r)&=&\frac{\sqrt{B}\left((n-2)A C^{-2+\frac{n}{2}} ((n-3)(n-2)-2 C(V+\Lambda))-C^{1-\frac{n}{2}} \left(4 (n-3) A^2 X_a^2+(n-2) \left(A X_t^2-X_\phi^2\right)\right)\right)}{2 \sqrt{A} (n-2) X_a}-{X_0\over X_a}.\label{BEA-1}\nonumber\\
 \end{eqnarray}
\end{subequations}
The remaining auxilliary functions $X_0, X_a, X_t,$ and $X_\phi$ can be determined from the Euler-Lagrange equations of BPS Lagrangian density $\mathcal{L}_{eff}$. The constraint equation for $B$ yields $X_0=0$ which further implies, from the constraint equation for $A_t$, ${dX_t\over dr}=0$, or $X_t=cx_t$ is a real constant. With these additional explicit functions of $X_0$ and $X_t$, the constraint equations for $A$ and $\phi$ implies, respectively,
\begin{eqnarray}
 {X_a}'(r)&=&-\frac{\sqrt{B} C^{1-\frac{n}{2}}}{2(n-2)A^{3/2}}\left((n-2) X_\phi^2-4(n-3) A^2 X_a^2\right),\label{CE Xa-1}\\
 {X_\phi}'(r)&=&-\sqrt{AB} C^{-1+\frac{n}{2}} V'(\phi),\label{CE Xphi-1}
\end{eqnarray}
These equations further implies the constraint equation for $C$ to be trivial.

\subsection{Tangherlini Black Holes}\label{Tangherlini BH}
In this case the action only contains the gravity and gauge fields such that we need to remove all terms in the effective and BPS Lagrangian densities that contain the scalar field. This can be done by taking $\phi$ and its potential $V$ to be zero which also imply $X_\phi=0$ such that the constraint equation \eqref{CE Xphi-1} is trivially satisfied and the constraint equation \eqref{CE Xa-1} is simplified to
\begin{equation}
 {dX_a \over dr}={2(n-3)\over (n-2)}\sqrt{A~B}~C^{1-{n\over 2}} X_a^2={(n-3)X_a\over 2C} {dC \over dr}
\end{equation}
whose solution is $X_a=cx_a C^{n-3\over 2}$, with $cx_a$ is the integration constant. Using additional explicit function of $X_a$ and comparing the Bogomol'nyi-like equations \eqref{BEAt-1} and \eqref{BEC-1}, we obtain
\begin{equation}
 A_t=c_t-{(n-2)\over 2(n-3)}{cx_t \over cx_a}C^{-{n-3\over 2}},
\end{equation}
with $c_t$ is the integration constant. Further comparing the Bogomol'nyi-like equations \eqref{BEA-1} and \eqref{BEC-1}, we may obtain a differential equation
\begin{equation}
 -8 cx_a^2 C{dA\over dC}=cx_t^2 (n-2)C^{3-n}+(n-3) \left(4Acx_a^2-(n-2)^2\right)+2(n-2) C\Lambda 
\end{equation}
whose solution is given by
\begin{equation}
 A=\frac{(n-2)}{4(n-3)cx_a^2}\left(\frac{cx_t^2}{C^{n-3}}+{(n-3)\over(n-1)} \left((n-2)(n-1)-2 C \Lambda\right)\right)+{c_a\over C^{n-3\over 2}},
\end{equation}
with $c_a$ is a real constant. As in the previous case, the solution for $B$ can be obtained from the Bogomol'nyi-like equation \eqref{BEC-1},
\begin{equation}
 B={(n-2)^2\over 16cx_a^2}{1\over A~C}C'(r)^2.
\end{equation}
If we take $C(r)=r^2$ and set $cx_a={(n-2)\over2},cx_t=-Q,$ and $c_a=-2M$ then we get the black hole type solutions,
\begin{eqnarray}
 ds^2&=&-\left(1-{2M\over C^{n-3\over 2}}+\frac{Q^2}{(n-3)(n-2)C^{n-3}}-{2C\Lambda\over(n-2)(n-1)}\right)dt^2+\left(1-{2M\over C^{n-3\over 2}}+\frac{Q^2}{(n-3)(n-2)C^{n-3}}-{2C\Lambda\over(n-2)(n-1)}\right)^{-1}dr^2\nonumber\\
 &&+r^2 d\Omega_{n-2}^2,
\end{eqnarray}
where $M$ is the black hole mass and $Q$ is the electric charge.

\subsection{Three dimensional $V$-scalar vacuum}\label{scalar hair BH}
Here $n=3$ without gauge field, $A_t=X_t=0$, such that the Bogomol'nyi-like equations for $\phi,C,$ and $A$ are, respectively,
\begin{subequations}\label{BE Electrovac-3}
 \begin{eqnarray}
  \phi'(r)&=&-\sqrt{B\over A~C} X_\phi,\label{BEph-3}\\
  C'(r)&=&4\sqrt{A~ B~C}~ X_a,\label{BEC-3}\\
  A'(r)&=&\sqrt{B\over A~C}\frac{1}{2 X_a}\left(X_\phi^2-2A C(V+\Lambda)\right),\label{BEA-3}\nonumber\\
 \end{eqnarray}
\end{subequations}
 and the constraint equations for $A$ and $\phi$, respectively,
\begin{eqnarray}
 {X_a}'(r)&=&-\frac{1}{2A^{3/2}}\sqrt{B\over C}X_\phi^2,\label{CE Xa-2}\\
 {X_\phi}'(r)&=&-\sqrt{A~B~C}~ V'(\phi).\label{CE Xphi-2}
\end{eqnarray}
As an example we fix the functions $C=r^2$ and $\phi={c_\phi\over r^p}$, with $p>0$ and $c_\phi$ is a real constant, such that $X_a={1\over 2\sqrt{AB}}$ and $X_\phi=p~\phi\sqrt{A\over B}$. Using these additional explicit functions of $X_a$ and $X_\phi$, the constraint equation \eqref{CE Xa-2} can be rewritten as 
\begin{equation}
 B'(r)=\frac{B}{\sqrt{C}}\left(2 B C (V+\Lambda)+p^2\phi ^2\right).\label{CE B}
\end{equation}
Furthermore we find that
\begin{equation}
    {d\log(A~B)\over dr}=-2p\phi {d\phi\over dr}
\end{equation}
whose solution is $A={c_a\over B} e^{-p\phi^2}$, with $c_a$ is an integration constant. Without loos of generality we can set $c_a=1$. On the other hand the constraint equation \eqref{CE Xphi-2} can also be rewritten, by using $V'(\phi)={V'(r)\over\phi'(r)}$, as
\begin{equation}
V'(r)=-{p^2\phi^2\over\sqrt{C}}\left({p\over B~C}+2(V+\Lambda)\right).\label{CE V}
\end{equation}
What remain to be solved are the constraint equations \eqref{CE B} and \eqref{CE V} whose solutions for $p=1$ are given by,
\begin{eqnarray}
 B(r)&=&\frac{c_b}{r^2 e^{\frac{c_\phi^2}{2 r^2}}\left(1+c_\phi^2 c_v e^{\frac{c_\phi^2}{2r^2}}\right)},\\
 A(r)&=&\frac{r^2}{c_b} e^{-\frac{c_\phi^2}{2 r^2}}\left(1+c_\phi^2 c_v e^{\frac{c_\phi^2}{2r^2}}\right),\\
 V(r)&=&-\Lambda-{e^{\frac{c_\phi^2}{2 r^2}}\over c_b}\left(1+c_\phi^2 c_v e^{\frac{c_\phi^2}{2 r^2}}\right) +\frac{c_\phi^4 c_v }{2 c_b}{e^{\frac{c_\phi^2}{r^2}}\over r^2},
\end{eqnarray}
with $c_b>0$ and $c_v<0$ are real constants. Here black hole solutions exist if $c_\phi^2c_v>-1$, with the effective cosmological constant $\lambda=-{1+c_\phi^2c_v\over c_b}$, and hence there are only asymptotically Anti-de Sitter black holes. These solutions are equal to solutions found in~\cite{Karakasis:2023ljt}, for $\Lambda=0$, and has been discussed there in great detail. For $p=2$ and $\Lambda=0$, the exact solutions are given by
\begin{eqnarray}\label{New BH2}
 B(r)&=&\frac{c_b e^{-\frac{2 c_\phi^2}{r^4}}}{r^2 \left(-\sqrt{\pi }~ \text{erf}\left(\frac{c_\phi}{r^2}\right)+4 c_v c_\phi\right)},\\
 A(r)&=&\frac{r^2}{c_b}\left(-\sqrt{\pi }~ \text{erf}\left(\frac{c_\phi}{r^2}\right)+4 c_v c_\phi\right),\\
 V(r)&=&-{2 c_\phi\over c_b r^2} e^{\frac{c_\phi^2}{r^4}}+{e^{\frac{2 c_\phi^2}{r^4}}\over c_b r^4} \left(2 c_\phi^2-r^4\right) \left(-\sqrt{\pi }~ \text{erf}\left(\frac{c_\phi}{r^2}\right)+4 c_v c_\phi\right),
\end{eqnarray}
where $c_v>0$ and $c_b\neq0$ are real constants, with $\mbox{sign}(c_b)=\mbox{sign}(c_\phi)$. The black holes exist if  $4c_v|c_\phi|<\sqrt{\pi}$, with the effective cosmological constant $\lambda=-{4c_\phi c_v\over c_b}$, and thus again there are only asymptotically Anti-de Sitter black holes. Near the boundary, $r\to\infty$, the solutions behave as
\begin{eqnarray}
 A(r\to\infty)&\sim&-\lambda r^2 -{2 c_\phi\over c_b}+O\left(\left( 1\over r\right)^3\right),\\
 V(r\to\infty)&\sim& c_\phi \lambda -{8 c_\phi^4 \over 3 c_b r^6} + O\left(\left( 1\over r\right)^7\right).
\end{eqnarray}
\begin{figure}[H]
    \centering
    \includegraphics[width=0.8\textwidth]{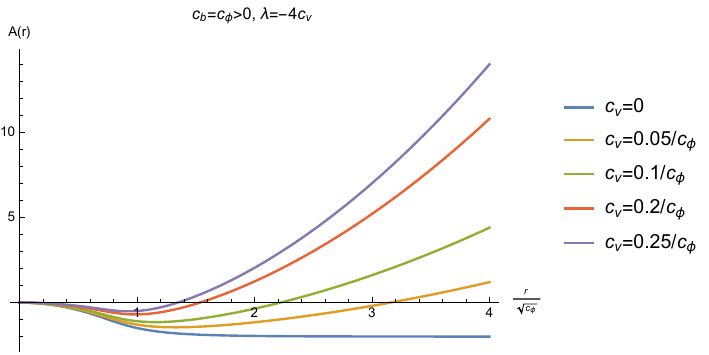}
    \caption{The horizon is determined by the value of $c_v c_\phi$, with fixed $c_b/c_\phi$. Its radius is larger as we increase $c_v c_\phi$.}
    \label{fig:An3}
\end{figure}
\begin{figure}[H]
    \centering
    \includegraphics[width=0.8\textwidth]{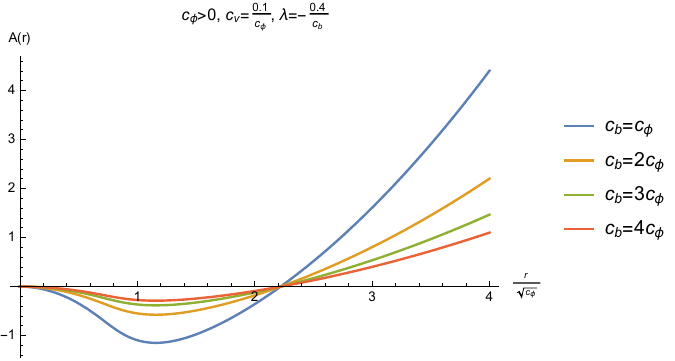}
    \caption{Here we can see the horizon radius does not change if we vary the value of $c_b/c_\phi$. For $c_v=0.1/c_\phi$, the horizon radius  is $r_h\approx2.22088\sqrt{c_\phi}$.}
    \label{fig:An3-1}
\end{figure}

\section{Remarks}
The first important step in the BPS Lagrangian method is to determine the effective Lagrangian density. For general solutions, without a priori imposing any ansatz, the effective Lagrangian density is simply its original Lagrangian density multipliying with the square root of determinant of the metric tensor. However in many cases, such as discussed in this article, it is simpler to work on solutions under particular ansatz direclty. In these cases we need to identify the effective Lagrangian density and its effective fields such that its Euler-Lagrange equations, for the effective fields, are effectively equal to the original Euler-Lagrange equations for the fundamental fields written under the corresponding ansatz. For the cases considered here the effective Lagrangian density for gravity parts are simply obtained by imposing the ansatz \eqref{ansatz}, after identifying the effective fields, into the action \eqref{EH action}. Knowing the correct effective fields is important for deriving the constraint equations from the BPS Lagrangian density.

In the case of static spherically symmetric vacuum, in section \eqref{uniqueness solution 1}, and of electrovac, in section \eqref{uniqueness solution 2}, both solutions for $B$, \eqref{sol B 1} and \eqref{sol B 2} respectively, are given by the same function of $C(r)$ and $C'(r)$ while the remaining functions can be written as functions of $C$. Therefore once the function $C$ is fixed the other functions can be determined completely. Taking $C=r^2$ gives us the Schwarzschild and Reissner-Nordstr\"{o}m black holes, and hence other solutions of $C$ are related to those black holes. Here by using the BPS Lagrangian method we have shown a simple alternative proof of the black hole uniqueness theorems~\cite{Israel:1967wq,Israel:1967za}. A rather complete proof of those black hole uniqueness theorems would be done by considering all possible terms in the BPS Lagrangian density, $\mathcal{L}_{BPS}$. In section \eqref{Tangherlini BH} we found that static spherically symmetric electrovac in higher dimensions $n>4$ are completely solved in terms of functions $C(r)$ and $C'(r)$, and hence other solutions are related to the Tangherlini black holes in which we fixed $C=r^2$. Here again we may have shown a simple (incomplete) proof of the black hole uniqueness theorem in higher dimensions as proposed in~\cite{Hollands:2012xy}. On the other hand, the solutions for $V-$scalar vacuum in section \eqref{ES gravity} can not be completely determined by functions $C(r)$ and $C'(r)$. Even after we fixed $C=r^2$ and the scalar field $\phi\equiv\phi(r)$, there were stil remaining first-order differential equations \eqref{Coulomb scalar} that need to be solved. One can easily check that the Schwarzschild black hole, with $B=\left(1-{r^2\over 3}\Lambda-{2M\over r}\right)^{-1}$, is not solutions to those first-order differential equations. This may show a violation of the black hole``no-scalar-hair'' theorem formulated in~\cite{Bekenstein:1995un}. The existance of these scalar hair black holes have been studied numerically for asymptotically flat metric, with $\Lambda=0$, in~\cite{Nucamendi:1995ex,Chew:2022enh,Chew:2023olq}. In section \eqref{ES gravity} we indeed found exact solutions of these scalar hair black holes given in \eqref{New BH1}. Although the scalar potential has similar profile, with negative value in whole space, as asymptotically AdS-black holes found in~\cite{Gonzalez:2013aca}, we nevertheless able to show that there exist asymptotically flat black holes in the Einstein-Scalar gravity~\eqref{action ES gravity}. We also found exact solutions \eqref{New BH2} of scalar hair black holes in three dimensional spacetime other then the one found in~\cite{Karakasis:2023ljt} for $n=2$.

In this article, we only considered the BPS Lagrangian density that are linear in first-order derivative of the effective fields. The most general BPS Lagrangian density that we can consider, e.g. for the effective Lagrangian density \eqref{eff Lagrangian ES gravity}, is
\begin{equation}
 \mathcal{L}_{BPS}=\sum^2_{\substack{i=0\\j+k+l+m=i}} X_{ijklm}(A,B,C) ~A'(r)^j B'(r)^k C'(r)^l\phi'(r)^m 
\end{equation}
that would allow us to completing the square $\mathcal{L}_{eff}-\mathcal{L}_{BPS}$ in $A'(r),B'(r),C'(r),$ and $\phi'(r)$. However the maximum auxilliary functions $X_{ijkl}$ that can be considered is five otherwise the set of equations could be underdetermined since there are only four constraint equations and one equation from the remaining terms in $\mathcal{L}_{eff}-\mathcal{L}_{BPS}$ after competing the square. Using this general BPS Lagrangian density, we may able to obtain a relation for functions $A$ and $B$, that is $A=1/B$, which is usually assumed in some of literatures on scalar hair black holes~\cite{Herdeiro:2015waa}.

\begin{acknowledgments}
We would like to acknowledge support from the ICTP through the Associates Programme (2018-2023).
\end{acknowledgments}

\bibliography{references}
\end{document}